\documentclass[pre,aps,twocolumn,showpacs,groupedaddress,superscriptaddress,amsmath,amssymb,10pt]{revtex4}
\usepackage{subfigure}
\usepackage{url,graphicx, color}
\usepackage{dcolumn}
\usepackage{bm}
\usepackage{epsfig,amsmath}
\usepackage{amssymb,bm}

\begin{document}

\title{A criterion for condensation in kinetically constrained one-dimensional transport models.}

\author{D.~M.~Miedema}
\email{D.M.Miedema@uva.nl}
\affiliation{Institute of Physics, University of Amsterdam, P.O.~box 94485, 1090 GL Amsterdam, The Netherlands}
\author{A.~S.~de Wijn}
\email{astrid@dewijn.eu}
\affiliation{Department of Physics, Stockholm University, 10691 Stockholm, Sweden}
\author{P.~Schall}
\email{ps@peterschall.de}
\affiliation{Institute of Physics, University of Amsterdam, P.O.~box 94485, 1090 GL Amsterdam, The Netherlands}

\begin{abstract}
We study condensation in one-dimensional transport models with a kinetic constraint. The kinetic constraint results in clustering of immobile vehicles; these clusters can grow to macroscopic condensates, indicating the onset of dynamic phase separation between free flowing and arrested traffic. We investigate analytically the conditions under which this occurs, and derive a necessary and sufficient criterion for phase separation. This criterion is applied to the well-known Nagel-Schreckenberg model of traffic flow to analytically investigate the existence of dynamic condensates. We find that true condensates occur only when acceleration out of jammed traffic happens in a single time step, in the limit of strong overbraking. Our predictions are further verified with simulation results on the growth of arrested clusters. These results provide analytic understanding of dynamic arrest and dynamic phase separation in one-dimensional traffic and transport models.
\end{abstract}

\pacs{64.70.qj, 64.70.P-, 64.70.Q-, 89.40.Bb, 64.75.Gh}{}

\maketitle

\section{Introduction}
A wide range of driven many-particle systems including traffic flow \cite{Chowdhury_review}, active colloids \cite{bartolo} and shaken granular gases \cite{granular_gas} exhibit interesting collective large-scale phenomena. The interactions between the driven constituent particles cause collective behavior such as collective slowing down that can eventually emerge into macroscopic phenomena affecting major parts of the system. In traffic, the interactions of vehicles that avoid collisions lead to nontrivial, strongly nonlinear flow behavior: increasing the number of vehicles does not necessarily result in an increase of the throughput. To understand and predict this behavior, it is important to know how the dynamics organize in space. Do all particles slow down gradually or do only certain particles slow down while others still move? An intriguing question in this context is whether the interactions between cars lead to macroscopic separation into arrested and moving traffic, a transition analogous to an equilibrium phase transition. The accumulation of a macroscopic fraction of all particles into a cluster is called condensation in real space. Recently, such condensation phenomena have been studied in one-dimensional transport models~\cite{PFSS,con_drift,ZRP_con}. Whether and how dynamic condensation occurs remains a largely open question and can only be addressed analytically in certain exactly solvable models. Most work has focussed on the exactly solvable zero-range process ZRP \cite{ZRP,ZRP_review} and related models \cite{chip1,chip2}. In these models, the dynamics of particles is typically specified per lattice site: particles can accumulate on a given site while hopping from one site to the next. It has been shown analytically that these systems exhibit condensation and symmetry breaking, even in one dimension.
The situation is, however, different in traffic models, where vehicles must follow each other and cannot accumulate on any site. In these models, the vehicle dynamics are set by a kinetic constraint between neighboring particles that guarantees vehicles do not collide. The question is whether in these models, condensate transitions can still occur, and how they can be analytically predicted based on the microscopic interactions of the vehicles or particles. Although condensation and phase separation phenomena have been investigated numerically in commonly used traffic models~\cite{Chowdhury_review,phase_sep_num} and some attempts have been made to connect traffic models with the ZRP \cite{traffic_zrp} and related models \cite{traffic_chipping1}, a general analytic treatment is lacking so far.

Here, we present just such an analytic criterion of condensate formation in traffic models. Our analysis is based on rate equations for the generation, growth, shrinkage and split-up of traffic jams. We establish a criterion for the occurrence of extensively large jams as a function of the microscopic interactions of the traffic model. We then apply this criterion to the well-known Nagel-Schreckenberg (NS) \cite{NS} model and related models of traffic flow to investigate analytically if and how condensation occurs, hence answering the longstanding question of dynamic condensates in this model~\cite{lifetimes}. We find that a true condensate forms only in the special case where the acceleration out of jammed traffic occurs in a single acceleration step, in the limit of infinite braking probability. In all other cases, arrested clusters either dissolve or split up. We illustrate the growth dynamics of these arrested clusters using long simulation runs, and demonstrate the similarity to the dynamics of a diffusion process. These simulation results validate our analytic criterion, and elucidate the formation kinetics of condensates.

\section{Criterion} \label{sec:crit}

\subsection{Class of Models} \label{sec:defmod}

We consider kinetically constrained one-dimensional transport models that are defined by mass-conserving local dynamical rules. The vehicles move unidirectionally over a discrete lattice in discrete time. A vehicle can move freely when it is out of the interaction range of other vehicles, but becomes kinetically constrained when it closely approaches another vehicle: it must slow down to avoid a collision. This kind of constraint is also present in more general models of transport of particles. The kinetic constraint can, for example, be a hard-core repulsion between neighboring particles. ``Softer'' constraints with longer range are also possible as long as the order of the particles is conserved. Once the kinetic constraint is released, particles accelerate with a certain probability back to free flow. We call the free flowing particles ``active", and the kinetically constrained particles ``inactive"


\begin{figure}
\epsfig{figure=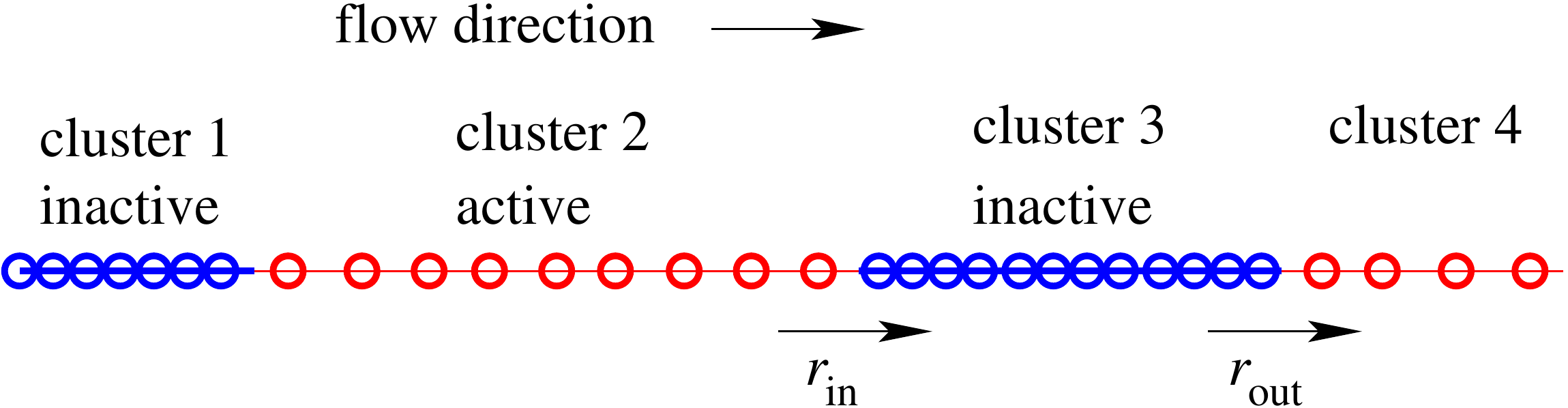,width=8.6cm}
\caption{
Schematic of vehicle transport in mass transport models. The dynamic interactions lead to clustering of immobile (inactive) particles. These inactive clusters coexist with clusters of mobile (active) particles. A few coexisting clusters are shown, as well as the inflow and outflow rate $r_{in}$ and $r_{out}$ of particles at the upstream and downstream boundary of an inactive cluster. The average distance between particles in active clusters must be larger than the interaction distance, while the distance between inactive particles is typically smaller than the interaction distance.
}
\label{fig:sketch}
\end{figure}

\subsection{Derivation of Criterion}
Inactive particles form clusters due to their dynamic interaction; here, we define a cluster as a sequence of particles in the same state (active or inactive). A typical particle configuration consists of several coexisting active and inactive clusters as shown in Fig.~\ref{fig:sketch}. These clusters can grow and eventually reach macroscopic size. We call a cluster a condensate, if in the limit of infinite system size, the cluster contains an infinite number of particles. Here, we allow short-lived interruptions in the sequence of inactive particles that exist on a timescale much shorter than the typical timescale of growth or shrinkage of the cluster. These ``bubbles'' disappear fast and almost always consist of only a single active particle.

There are several competing processes that lead to growth or shrinkage of clusters. Clusters can grow one by one by vehicles leaving or entering at the
boundaries (see cluster 3 in Fig.~\ref{fig:sketch}). Clusters can also split up into two by vehicles changing their state inside a cluster. Finally, two clusters can merge when the cluster that separates them shrinks to zero.

Below we will analyze these cluster processes in detail to find the condition for condensation. The idea is as follows:
1. Inactive clusters must be unable to split up in order to become infinitely large.
2. Inactive clusters must grow, i.e. their growth rate must be at least as large as their shrink rate. The growth of existing clusters is, however, reduced by any new inactive cluster that forms upstream; such new inactive cluster takes up particles and reduces the inflow of existing downstream inactive clusters.

The two conditions described above have to be met independently: because the split-up rate of clusters scales with the cluster size, while the growth rate of clusters does not (it is always limited to maximum 1 particle per timestep), the two processes cannot balance, and both conditions must be fulfilled simultaneously.

\subsubsection{Splitting up of inactive clusters}
We first investigate the split-up of inactive clusters. Split-up occurs when the distance between inactive vehicles increases spontaneously releasing the kinetic constraint. Such split-up is detrimental for condensation. To prevent it, density fluctuations inside the cluster should not occur, i.e. the density $\rho_\mathrm{ina}$ inside the inactive cluster should be maximal, $\rho_\mathrm{ina} = \rho_\mathrm{max}$. For hard-core repulsion, we have $\rho_\mathrm{max}=1 ~$\footnote{It is possible to construct systems that have a range of densities that make fluctuations impossible.  In this case, by $\rho_\mathrm{max}$ we mean any density in this range.}.

If $\rho_\mathrm{ina}$ is lower than $\rho_\mathrm{max}$, density fluctuations do exist; nevertheless, condensation will still occur if these density fluctuations are short-lived. This is the case when the density in inactive regions is much larger than that of active ones, i.e. when $\rho_\mathrm{act}/\rho_\mathrm{ina} \rightarrow 0$. In this case, any active ``bubble" requires an infinite amount of space; that much space is not available inside inactive clusters, and such temporal bubble will disappears almost immediately.
We thus obtain the first condition for condensation:
\begin{align}
\rho_\mathrm{ina} = \rho_\mathrm{max} ~~~ \vee ~~~ \frac{\rho_\mathrm{act}}{\rho_\mathrm{ina}} \rightarrow 0~.
\label{eq:crit2}
\end{align}

We note that this condition also implies that inactive clusters cannot merge.

\subsubsection{Growth versus creation of inactive clusters}
We now consider the processes that grow and shrink the inactive cluster due to in- and outflow of single vehicles. An inactive cluster grows due to vehicles entering at the upstream boundary at rate $r_\mathrm{in}$, while it shrinks due to vehicles leaving the cluster at the downstream boundary with rate $r_\mathrm{out}$, see cluster 3 in Fig.~\ref{fig:sketch}. These two processes grow and shrink the inactive cluster, respectively, with rates $r_+$ and $r_-$ according to:
\begin{align}
\label{eq:ratesplus}
r_+ &= r_\mathrm{in} (1-r_\mathrm{out})~, \\
r_- &= (1-r_\mathrm{in}) r_\mathrm{out}~.
\label{eq:ratesmin}
\end{align}
A cluster can only grow persistently if $r_+ \geq r_-$; only then, we expect a condensate to form. Since in steady-state, $r_+ > r_-$ is not possible due to particle conservation, this leaves us with the condition $r_+ = r_-$. This means, for condensation to occur, the difference $\Delta r=r_- - r_+~$ must vanish relative to the absolute value of $r_+$ or $r_-$ that sets the typical time scale of the system. Hence
\begin{align}
\frac{r_- - r_+}{r_-} = \frac{\Delta r}{r_-} \rightarrow 0~.
\label{eq:crit2pre}
\end{align}
The task is now to find an expression for $\Delta r$ in terms of basic dynamical quantities. We rewrite $\Delta r$ using Eqs.~(\ref{eq:ratesplus}) and (\ref{eq:ratesmin}) to relate it to the in- and outflow rate of vehicles,
\begin{align}
\Delta r = (1-r_\mathrm{in}) r_\mathrm{out}-r_\mathrm{in} (1-r_\mathrm{out}) =  r_\mathrm{out} - r_\mathrm{in}.
\label{eq:deltar2}
\end{align}
Here, the inflow rate $r_{in}$ of the inactive cluster depends on the velocity $\tilde{v}$ of vehicles leaving the upstream active cluster, and on their density, $\tilde{\rho}$. One can easily see that
\begin{align}
r_\mathrm{in}=\tilde{\rho}(\tilde{v}-v_\mathrm{c}),
\label{eq:r_in}
\end{align}
where, $v_\mathrm{c}$ is the average (negative) velocity at which the cluster boundary travels backwards. Both $\tilde{\rho}$ and $\tilde{v}$ themselves depend on processes in the active cluster 2: they become reduced when new inactive clusters form inside the active cluster. Likewise, this depends on the density in region 2, which itself is controlled by the outflow of the upstream inactive cluster 1. For the density in the active cluster 2, we thus find the upper bound
\begin{align}
\tilde{\rho} = \frac{r_\mathrm{out}}{v_\mathrm{act}-v_\mathrm{c}},
\label{eq:densIN}
\end{align}
provided that $\tilde{\rho}$ becomes larger than the global density.
To obtain an expression for the velocity $\tilde{v}$, we introduce the fraction $\tilde{f}$ of inactive particles in region 2. We can then write
\begin{align}
\tilde{v} = \tilde{f} v_\mathrm{ina}+(1-\tilde{f}) v_\mathrm{act} ~.
\label{eq:vIN}
\end{align}
By inserting Eqs.~(\ref{eq:r_in}) - (\ref{eq:vIN}) in Eq.~(\ref{eq:deltar2}), we find that
\begin{align}
\Delta r &=r_\mathrm{out}-\frac{r_\mathrm{out}}{v_\mathrm{act}-v_\mathrm{c}} \lbrack \tilde{f} (v_\mathrm{ina}-v_\mathrm{c})+(1-\tilde{f})(v_\mathrm{act}-v_\mathrm{c})\rbrack ~,\\
&=r_\mathrm{out}\tilde{f} \frac{v_\mathrm{act}-v_\mathrm{ina}}{v_\mathrm{act}-v_\mathrm{c}} ~,
	\label{eq:deltar}
\end{align}
which relates $\Delta r$ to the car velocities and outflow rates. Finally, we express the fraction $\tilde{f}$ of inactive particles in terms of the creation rate $u$ per particle of inactive clusters, their average lifetime, $T$, and their average length, $n$. In steady-state this fraction is
\begin{align}
\tilde{f} &= u T n~.
\label{eq:fI}
\end{align}
Using Eqs.~(\ref{eq:deltar}) and~(\ref{eq:fI}), our criterion for the growth rate of clusters [Eq.~(\ref{eq:crit2pre})] then becomes
\begin{align}
\frac{\Delta r}{r_-} = \frac{r_\mathrm{out} u T n ( v_\mathrm{act}  -  v_\mathrm{ina} )} {r_- ( v_\mathrm{act}  -  v_\mathrm{c} )} \rightarrow 0~,
	\label{eq:crit01}
\end{align}
which simplifies to
\begin{align}
\frac{r_\mathrm{out} u T n} {r_- } \rightarrow 0~,
	\label{eq:crit1}
\end{align}
because $v_\mathrm{act}>v_\mathrm{ina}$ and $v_\mathrm{c}<0$, and all velocities are finite. Eq.~(\ref{eq:crit1}) provides the second criterion for condensation. It ensures that the growth rate of inactive clusters is at least as large as their shrink rate, so that inactive clusters can grow \footnote{We note that in order to derive Eq.~(\ref{eq:crit1}), we assumed high density. For $\rho<\frac{r_\mathrm{out}}{v_\mathrm{act}-v_\mathrm{c}}$, it follows from Eqs.~(\ref{eq:densIN}) and~(\ref{eq:vIN}) that $r_\mathrm{in}<r_\mathrm{out}$.}.

We thus arrive at a twofold criterion for condensate formation, consisting of Eqs.~(\ref{eq:crit2}) and~(\ref{eq:crit1}).
The first equation guarantees that inactive clusters do not split up; the second equation assures that the growth of inactive clusters is not hindered by the formation of new inactive clusters. Together, these two equations provide a necessary and sufficient condition for condensation.

\section{Application to traffic model}

We now apply the criterion, Eqs.~(\ref{eq:crit2}) and~(\ref{eq:crit1}), to specific traffic models to demonstrate the occurrence or absence of dynamic condensates. In particular, we focus on the Nagel-Schreckenberg model of traffic flow~\cite{NS}, a well-studied simple model that captures much of the behavior of real traffic.

\subsection{Nagel-Schreckenberg model\label{sec:NS}}

The NS-model is a one-dimensional cellular automaton model with discrete time and space. The road consists of a regular lattice of $L$ sites, occupied by $N$ cars with average density $\rho = N / L$. Cars move with integer velocity over the lattice and are updated synchronously. The velocity $v_i$ of car $i$ can be at most the maximum velocity $v_{max}$, and becomes constrained when the distance to the next car $d_i < v_i$. The following dynamical update rules for the NS-model are applied in parallel to all $N$ cars:
\begin{itemize}
\item[1] \textit{Acceleration}: $v_i \rightarrow$ min($v_i+1,v_{max}$).
\item[2] \textit{Avoiding collisions}: If $d_i < v_i$ then $v_i=d_i$.
\item[3] \textit{Randomization}: Decrease $v_i$ obtained in the previous steps by 1, to a minimum of 0, with probability~$p$.
\item[4] \textit{Position update}: $x_i \rightarrow x_i+v_i$, $d_i \rightarrow d_i - v_i + v_{i+1}$.
\end{itemize}
The only source of stochasticity in the model is the randomization parameter $p$ that reflects the drivers' individual freedom to decelerate below $v_{max}$. We define car $i$ as freely flowing (active) if before the randomization step (3), $v_i=v_{max}$. While at low density, most cars move freely, at high density or large braking probability, $p$, jams (inactive clusters) form. The density above which stable jams form, was estimated by Gerwinski and Krug as $\rho_{tra}=(1-p)/(v_\mathrm{max}+1-2p)$ ~\cite{gerwinski}.

\begin{figure}[t]
\centering
\subfigure[]{
	\includegraphics[width=3cm]{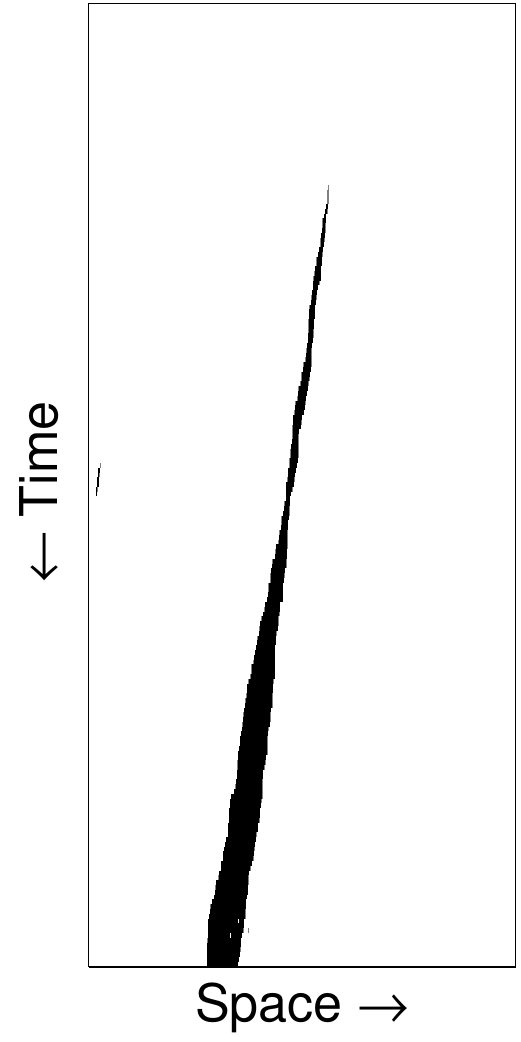}
	\label{fig:ns1_p1_snapshot}
}
\subfigure[]{
	\includegraphics[width=3cm]{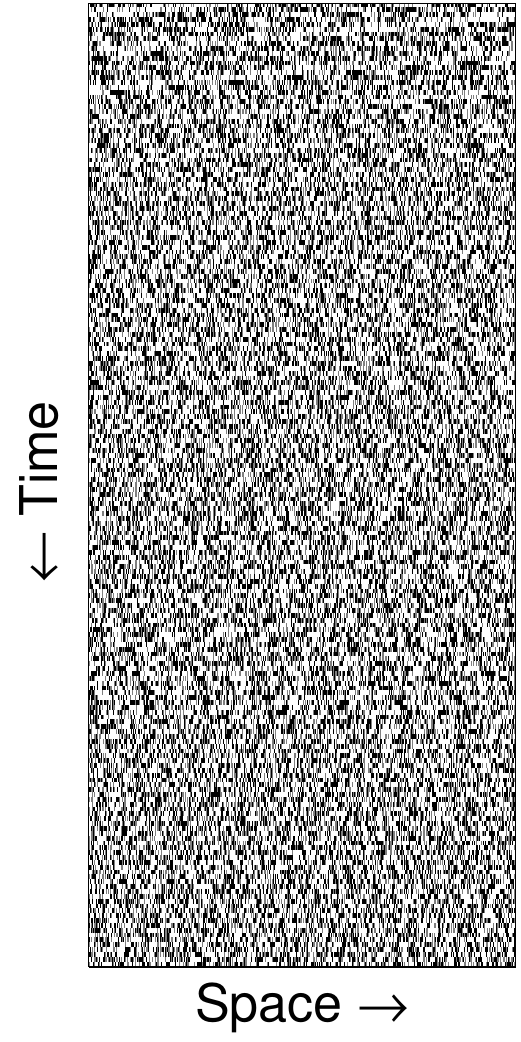}
	\label{fig:ns1_p0_snapshot}
}
\caption{
Space-time diagram of vehicles in the two deterministic limits of the NS-model, for $p=0.0002$ (a) and $p=0.9998$ (b). Inactive vehicles are indicated in black. The horizontal axis represents the car index. The simulations are performed at densities 20 \% above the transition density $\rho_\mathrm{tra}$.
}
\label{fig:snapshots}
\end{figure}

From simulations the idea has emerged that no sharp transition between free flow and jammed traffic occurs for finite stochasticity \cite{NSphasetransitionpolemic,Boccara,Chen}. This implies that there is no condensate. Condensates might, however, form in the deterministic limits $p \rightarrow 1$ and $p \rightarrow 0$ of the model. To appreciate the strikingly different behavior in the two deterministic limits, we show space-time diagrams constructed from simulations in Fig.~\ref{fig:snapshots}. In the limit $p \rightarrow 1$, a condensate forms as illustrated by the thick black line in Fig~\ref{fig:snapshots}a. A jam nucleates and grows into a condensate that contains all excess particles above the critical density. In contrast, in the limit $p \rightarrow 0$, there are many small jams (Fig.~\ref{fig:ns1_p0_snapshot}) that do not coalesce, and no macroscopic condensate forms. Some jams disappear and new jams are created. Below we investigate analytically the formation of condensates for different cases of $p$, starting with $v_\mathrm{max}=2$.
.

\subsection{Nagel-Schreckenberg with $0<p<1$ \\($v_{max}=2$)}
Simulations suggest that for finite stochasticity, $0<p<1$, there is no condensate. Indeed we will show that in this case, the second condition [Eq.~(\ref{eq:crit1})] is not fulfilled. To see this, we first note that for finite $p$, vehicles slow down randomly, and the average velocity is smaller than $v_{max}$. Hence, the inflow rate of jams, $r_\mathrm{in}$, is smaller than 1. Since we can rewrite $r_\mathrm{out}/r_-=1/(1-r_\mathrm{in})$ using Eq.~(\ref{eq:ratesmin}), we conclude that the first factor in Eq.~(\ref{eq:crit1}), $r_\mathrm{out}/r_- > 0$.

Furthermore, also $u>0$: due to velocity fluctuations at finite $p$, the distances between cars varies and cars can come within the interaction range with finite probability. Hence, new jams are formed even at arbitrarily low density and $u>0$.

Because the remaining factors in Eq.~(\ref{eq:crit1}), $T$ and $n$, are always larger than zero (a jam always exists for at least one time step and consists of at least one car), we conclude that Eq.~(\ref{eq:crit1}) is not fulfilled and thus there is no condensate. This is in agreement with the consensus in the literature about the absence of a sharp transition between free flowing and jammed traffic for $0<p<1$ \cite{NSphasetransitionpolemic,Boccara,Chen}.

\subsection{Nagel-Schreckenberg in the limit $p \rightarrow 1$ ($v_{max}=2$)}
In the limit $p \rightarrow 1$, cars almost always overbrake. To determine whether condensation occurs in this limit, we analyze the scaling of all quantities in Eqs.~(\ref{eq:crit2}) and~(\ref{eq:crit1}) as a function of the vanishing distance to the deterministic point: $\Delta p = 1-p \rightarrow 0$.

With $v_{max} = 2$ and $p \rightarrow 1$, free flowing traffic has average velocity $v_{max} - p = 1$, and jammed traffic has velocity 0. Hence, cars accelerate in a single step out of the inactive cluster, and the outflow rate equals the probability of acceleration, $r_\mathrm{out}=\Delta p$. According to Eq.~(\ref{eq:densIN}), it then follows that the density in active clusters scales as $\rho_\mathrm{act} \sim {\Delta p}$. Meanwhile, the density of a jam, $\rho_\mathrm{ina}$, is bounded from below due to the finite interaction range, and must be higher than $1/v_\mathrm{max}$. Consequently, the first part of the criterion, Eq.~(\ref{eq:crit2}), is fulfilled.

To check the second part, Eq.~(\ref{eq:crit1}), we note that because $r_{in} \rightarrow 0$ in the limit $p \rightarrow 1$, we can approximate $r_-=(1-r_\mathrm{in}) r_\mathrm{out} \approx r_\mathrm{out}$. We thus find that
\begin{align}
\frac{r_\mathrm{out} u T n}{r_-} \rightarrow u T n~.
\label{eq:critp1step1}
\end{align}
reducing the criterion to the scaling of $u$, $T$ and $n$.

\begin{figure}
\includegraphics[height=5.75cm]{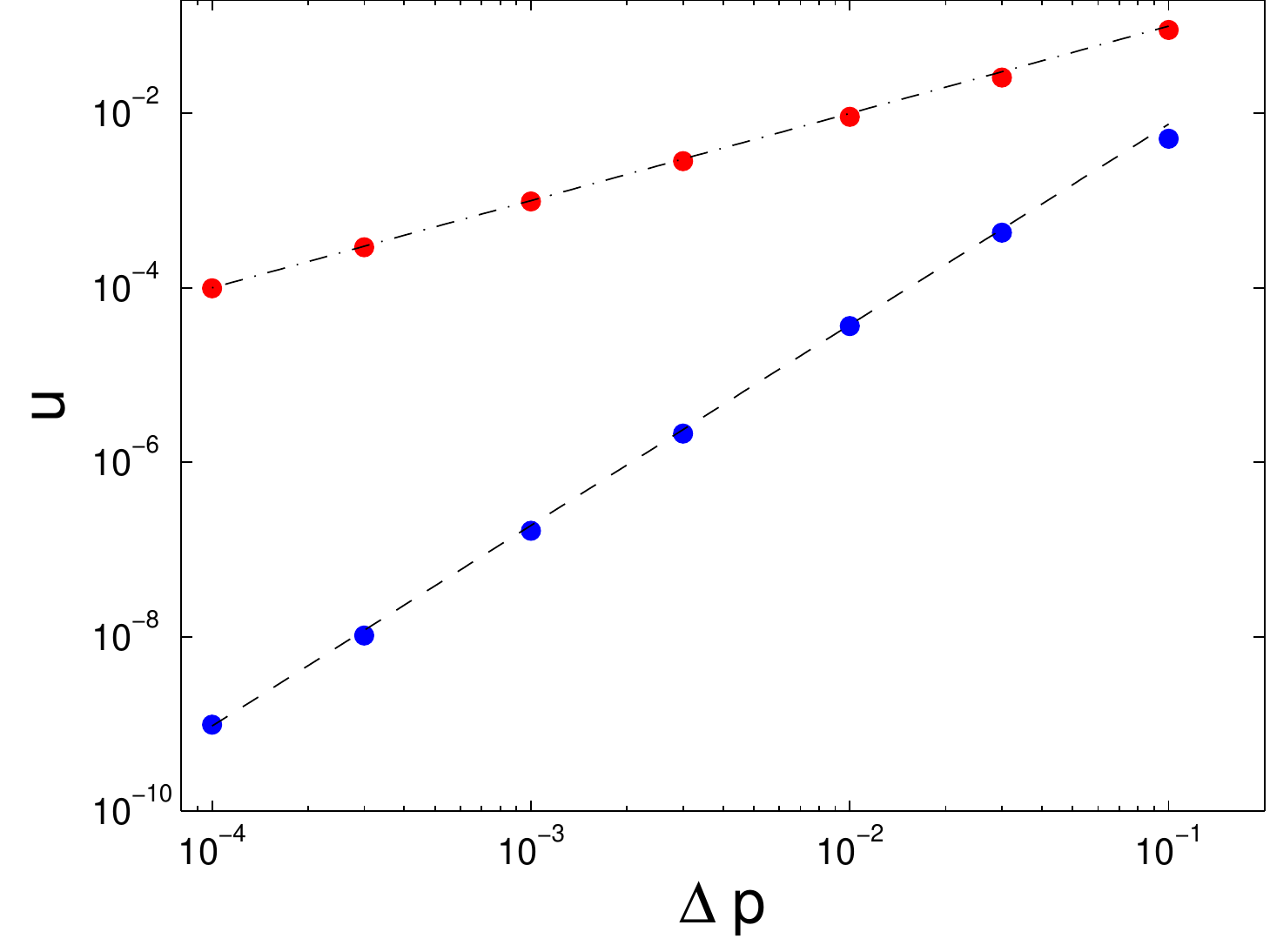}
\caption{Creation rate per car of new jams as a function of the distance $\Delta p$ to the deterministic point $p \rightarrow 0$ (red points, $\Delta p=p$) and $p \rightarrow 1$ (blue points, $\Delta p=1-p$). The dashed lines have slope 1 (red data points) and slope 2.3 (blue data points).}
\label{fig:ns_up}
\end{figure}

The scaling of $u$ can be estimated as follows: The distance between cars behaves as a diffusion process. Hence, we can estimate the creation rate $u$ of new jams from the time $\tau$ it takes for the root mean square of the change $\Delta d$ of the distance $d$ between subsequent cars to grow to the average distance itself: $\Delta d \approx \langle d \rangle$. For a random walker, the number of changes necessary to accumulate a change of $\langle d \rangle$ is $\langle d \rangle^2$, while for ballistic motion, the number of changes is $\langle d \rangle$. We will allow for a general power $\langle d \rangle^\beta$. Because the time to change the distance between two cars by one is of order $(\Delta p)^{-1}$, we obtain
\begin{align}
\tau \sim {\Delta p}^{-1-\beta}.
\label{eq:deltad2}
\end{align}
Because $u \approx 1/\tau$, we obtain $u \approx {\Delta p}^{\beta+1}$.
With simulations we find $u \sim {\Delta p}^{2.3 \pm 0.1}$ (Fig.~\ref{fig:ns_up}), and hence $\beta = 1.3$, an exponent between random walk and ballistic motion. The quantity $u$ thus vanishes on approach of the deterministic point.

For $n$, we find that maximally, its scaling is constant: Since $n$ indicates the number of cars in a jam, any persistent growth of $n$ would immediately imply that a condensate forms, and hence according to Eq.~(\ref{eq:crit1}) that its scaling must be bound. For $T$, i.e. the average lifetime of a jam, we estimate the scaling from the average time it takes for a car to accelerate out of a jam; this time diverges as $1/r_\mathrm{out}=1/\Delta p$. Hence, the average jam lifetime scales as $T \sim \mathcal{O}({\Delta p}^{-1})$ if $n$ is constant; any faster decrease would imply that the number of cars in a jam grows and thus again that a condensate forms.

With the scaling obtained for $u$, $n$ and $T$, Eq.~(\ref{eq:critp1step1}) becomes
\begin{align}
u T n \sim {\Delta p}^{\beta}~.
\end{align}
This quantity goes to zero in the limit $\Delta p \rightarrow 0$ for all $\beta$ values between the ballistic ($\beta=1$) and random walk ($\beta=2$) limit, thus meeting the requirement for condensate formation. We therefore expect a condensate to form, in agreement with the simulation results shown in Fig.~\ref{fig:snapshots}.

\begin{figure}
\includegraphics[height=6.5cm]{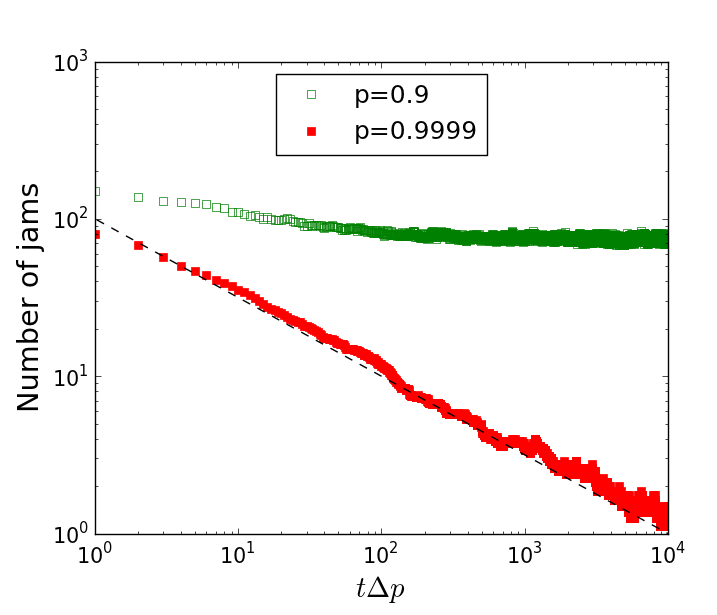}
\includegraphics[height=6.5cm]{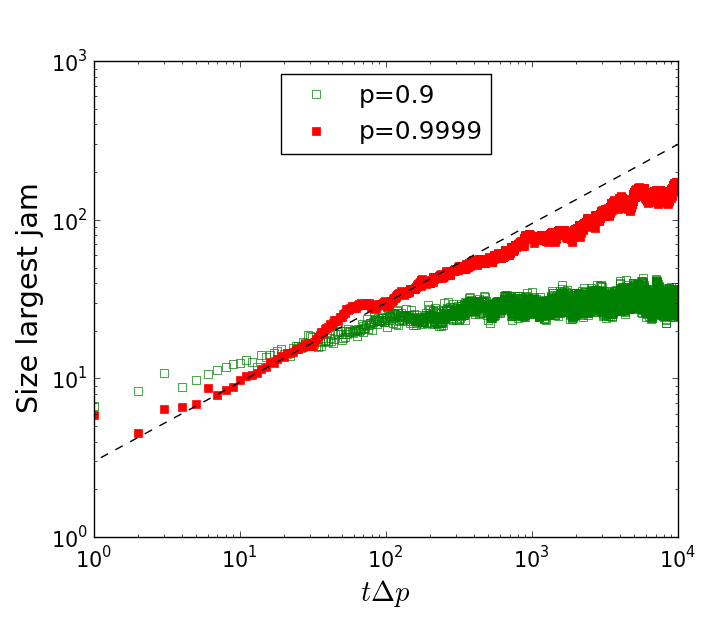}
\put(-220,205){(a)}
\put(-220,0){(b)}
\label{fig:ns1_nc}
\caption{Number of cars in the largest jam (a) and number of jams (b) versus time for p approaching 1. In the limit $p \rightarrow 1$, the jamsize diverges and the number of jams decreases over time indicating condensate formation. The dashed lines have a slope of $\pm$1/2.
}
\label{fig:ns1_nj}
\end{figure}

It is interesting to investigate the time dependence of the condensation process. In Fig.~\ref{fig:ns1_nj}, we plot the number of jams and the growth of the largest jam as a function of time. For $p$ close to 1, the number of cars in the largest jam increases with a power of $1/2$, while the number of jams decreases accordingly. This power-law scaling is reminiscent of the diffusive dynamics of the random-walk process, in which the probability of attachment of a car equals that of detachment. Indeed, we have shown above that a necessary criterion for condensation is $r_+=r_-$, i.e. inactive clusters increase or decrease with equal probability. This analogy between the size of jams and the position of a random walker was pointed out before by Nagel and Paczuski for the cruise control limit of the NS model \cite{NagelPaczuski}, and our analytical model predicts it as a necessary condition. We thus find that our criterion concludes correctly on the existence of dynamic condensates, and predicts the dynamics of their growth through a random-walk process.

\subsection{Nagel-Schreckenberg in the limit $p \rightarrow 0$ \\($v_{max}=2$)}
Simulations suggest that in this limit, no condensate forms ~\cite{NSp0transition1,NSp0transition2}. We will address this issue with the criterion starting with Eq.~(\ref{eq:crit1}). For $p \rightarrow 0$, the braking probability $p$ is vanishingly small. As a result, the outflow rate of jams $r_\mathrm{out}=1-p$. Using $r_\mathrm{in}=r_\mathrm{out}-\Delta r$, we can hence approximate $r_-=r_\mathrm{out} (1-r_\mathrm{in}) \approx p+\Delta r$. {\it A priori} we do not know which term dominates the scaling of $r_-$ when $p$ vanishes: $p$ or $\Delta r$. If $\Delta r$ determines the scaling, we immediately see that the left-hand side of Eq.~(\ref{eq:crit01}): $\Delta r/r_-=\Delta r/\Delta r \neq 0$ and there is no condensation. If $p$ determines the scaling, we can simplify Eq.~(\ref{eq:crit1}) as follows: Because the outflow rate is close to unity, the density of free flow is high and any random slow down of a car immediately causes the upstream neighbor to become kinetically constrained. Because this happens with probability $p$, the jam creation rate per car is $u \sim p$, as is also shown by the simulation results in Fig.~\ref{fig:ns_up}. With $u \sim p$, $r_- \sim p$ and $r_\mathrm{out} \approx 1$, Eq.~(\ref{eq:crit1}) becomes:
\begin{align}
\frac{r_\mathrm{out} u T n}{r_-} \sim T n~.
\label{eq:critp0}
\end{align}
Since both $T>0$ and $n>0$, we conclude that there is no condensation in the limit $p \rightarrow 0$, in agreement with the simulation results~\cite{NSp0transition1,NSp0transition2}.

\subsection{Nagel-Schreckenberg with $v_\mathrm{max}>2$}

It is frequently assumed that the NS model behaves qualitatively similar when changing $v_\mathrm{max}$~\cite{Schadschneider1999,vmax}. Here, we will investigate this analytically. Surprisingly, we find that for $v_\mathrm{max}>2$, there is no condensation in the limit $p \rightarrow 1$, in contrast to $v_\mathrm{max} = 2$. 

To see this, we note that in contrast to $v_\mathrm{max}=2$, where acceleration from jam to free flow occurs in a single step, for $v_\mathrm{max}>2$, the acceleration needs multiple steps. A car leaving the jam is still part of the jam until it reaches the maximum velocity. This lowers the density of jams, and leaves Eq.~(\ref{eq:crit2}) unfulfilled. In the spaces created inside the jam, new free flow can emerge that splits up the jam. This mechanism prevents the formation of an infinitely large jam.

\begin{figure}[t]
\centering
\subfigure[]{
	\includegraphics[angle=90, width=3cm]{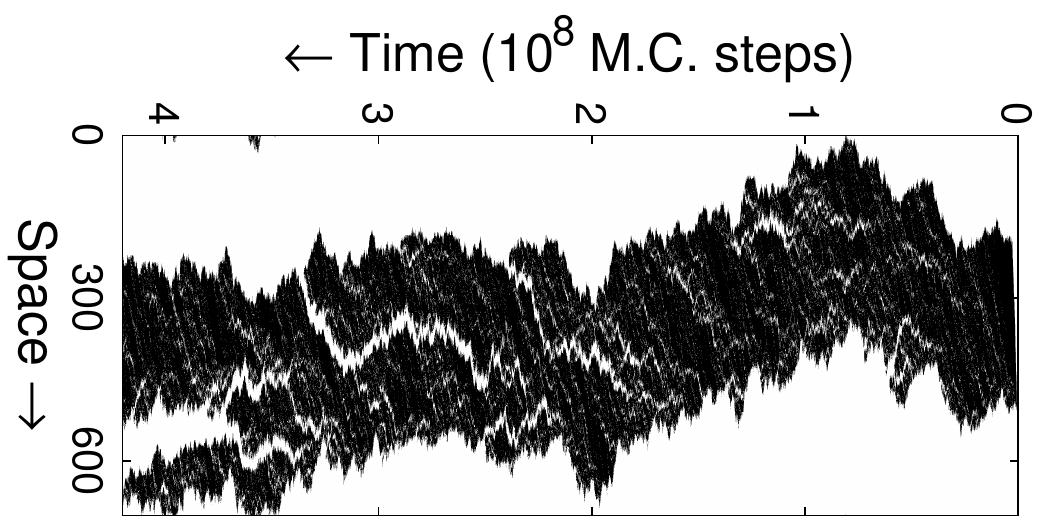}
	\label{fig:vm3_snapshot}
}
\subfigure[]{
	\includegraphics[angle=90, width=3cm]{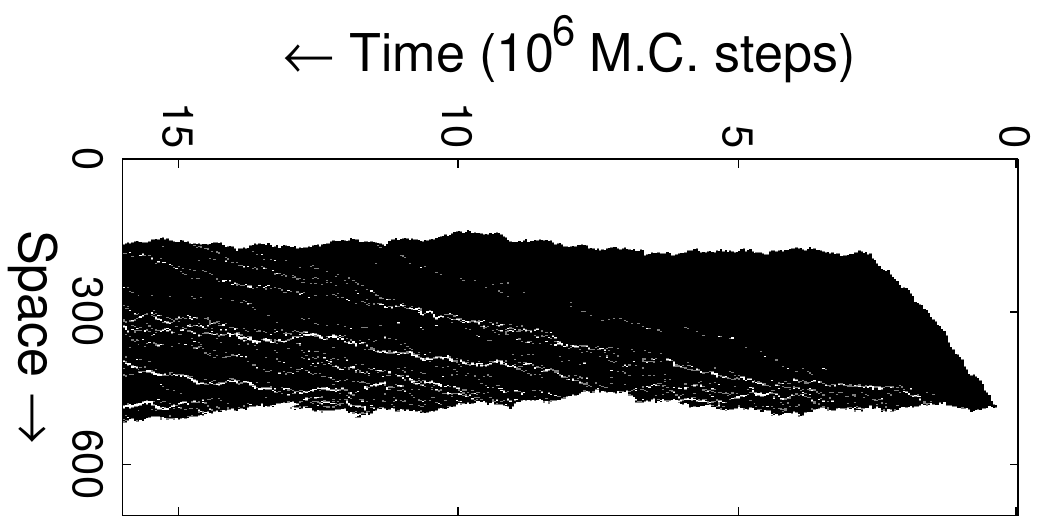}
	\label{fig:vm3_snapshot_zoom}
}
\caption{
Space-time diagram of vehicles in the NS-model with $p=0.9998$ and $v_\mathrm{max}=3$, entire simulation run (a), and enlarged section at early times (b). Inactive cars are shown in black. Horizontal axis represents car index; to follow the evolution of jams, we plot the space-time diagram in a frame co-moving with the speed of jams. The emerging white regions inside the jam indicate a split up of the original jam that becomes more pronounced at later times. The enlarged section in (b) shows that this split-up emerges at the downstream boundary of the jam. }
\label{fig:vm3_snapshot}
\end{figure}

We demonstrate the split up of jams in the space-time diagram obtained in simulations, see Fig.~\ref{fig:vm3_snapshot}. The simulation starts from random initial car positions; after a jam has nucleated, it grows, but shortly after that, the first free flow starts to appear inside the jam. This becomes most obvious in Fig.~\ref{fig:vm3_snapshot_zoom}, where we show a magnified section at early times. All free flow `bubbles' inside the jam clearly emerge at the downstream boundary of the jam. This free flow is persistent and covers larger regions at later times. These pictures demonstrate that there is no single macroscopic condensate for $v_\mathrm{max}>2$.

To complete the analytical discussion of the NS model we shortly comment on the limits $0<p<1$ and $p \rightarrow 0$ for $v_\mathrm{max}>2$. In both cases, the argument is similar to that of $v_\mathrm{max}=2$. For $0<p<1$, fluctuations in velocity create fluctuations in distances between free flowing cars. As a result, the creation rate of jams $u>0$. In the limit $p \rightarrow 0$ the creation rate of jams vanishes with $p$, $u \sim p$, but the growth rate of jams vanishes just as quickly, so there is no condensation.

In summary, the surprising conclusion of our analytical treatment of the NS model is that only in the case $v_\mathrm{max}=2$ (limit $p \rightarrow 1$) there is a true condensate transition.

\subsection{Application to Velocity Dependent Randomization model}
An extension of the Nagel-Schreckenberg model is the Velocity Dependent Randomization (VDR) model, in which there are \emph{two} fluctuation parameters instead of \emph{one}: $p_\mathrm{f}$ controls the fluctuations in free flow, while $p_\mathrm{j}$ controls the fluctuations of jammed traffic. This model takes account of the fact that drivers may behave differently depending on the traffic context, free flowing or jammed. We will show that in this model, where we have two control parameters, one for the creation rate and one for the growth rate of clusters, there is a condensate even in the limit $p_f,p_j \rightarrow 0$.

\begin{figure}
\includegraphics[height=5.75cm]{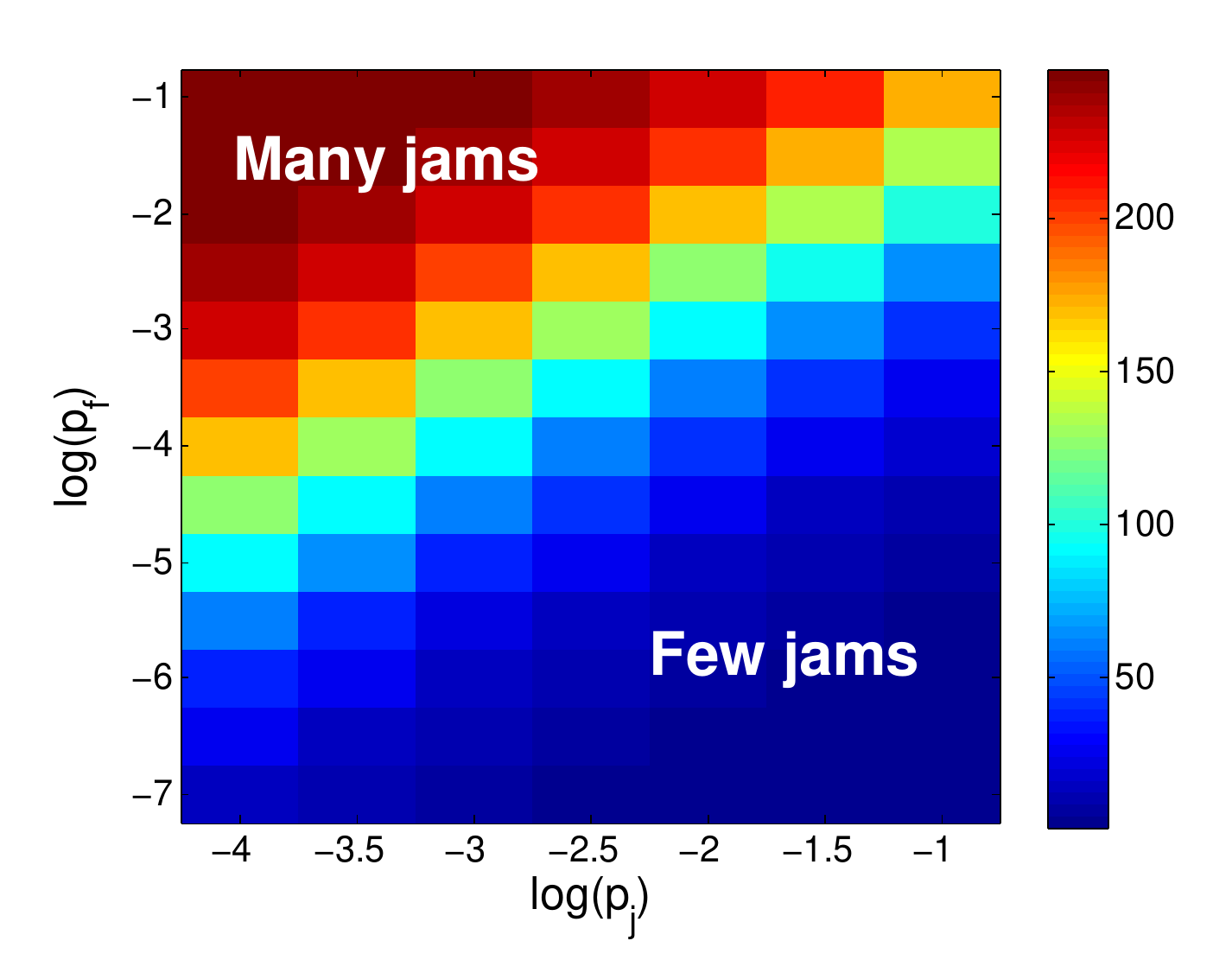}
\caption{Diagram illustrating the number of cars in the largest jam (see colorbar) as a function of $p_f$ and $p_j$ in the velocity-dependent randomization model. Simulations were performed with 1000 cars at density $\rho=0.4$.  The diagram is plotted in the region close to $p_f \rightarrow 0$ and $p_j \rightarrow 0$. Whether a condensate forms depends on how the limit to zero is taken. A few jams signal a condensate forms only in the limit $p_f / p_j \rightarrow 0$
}
\label{fig:vdr_nc}
\end{figure}

To do so, we use simulations to determine the size of the largest jam as a function of $p_j$ and $p_f$. To incorporate both $p_f$ and $p_j$, we modify the NS-model update scheme by adding an extra step before the randomization step 3. If car $i$ is jammed ($v_i<v_\mathrm{max}$ after step 2) then $p=p_j$, and if car $i$ is free flowing ($v_i=v_\mathrm{max}$) then $p=p_f$. Further, the update scheme is identical to the NS update scheme. We plot the size of the largest jam in a two-dimensional contour plot in Fig.~\ref{fig:vdr_nc}. This plot shows that a condensate forms if
\begin{align}
\frac{p_f}{{p_j}} \rightarrow 0~.
\label{eq:vdrcritfinal}
\end{align}
This numerical finding is indeed in line with the qualitative argument that the creation rate of new jams, controlled by $p_f$, must vanish faster than the growth rate of jams, controlled by $p_j$. This finding thus again demonstrates the principle of the criterion for condensation.

\section{Conclusion}
We have derived a criterion for condensation in one-dimensional transport models with a kinetic constraint that causes clustering of immobile particles. Whether this clustering leads to macroscopic phase separation depends on two factors: Firstly, density fluctuations in inactive clusters must be small enough to prevent the split-up of inactive clusters. Secondly, the growth rate of inactive clusters must dominate the formation rate of new inactive clusters since those reduce the inflow of existing clusters downstream.

The latter condition also means that condensation is only possible if the growth rate of inactive clusters equals their shrink rate. This establishes a generic analogy of the size of inactive clusters to the position of a random walker, that was previously found by Nagel and Paczuski \cite{NagelPaczuski} and Barlovic {\it et al.} \cite{Barlovic} for specific models. With this analogy we can explain the growth dynamics of condensates as well as the distribution of lifetimes and sizes of inactive clusters upon condensation.

We have applied the criterion to the well-known Nagel-Schreckenberg traffic model and find that a condensation transition occurs only in the limit $p \rightarrow 1$ with $v_\mathrm{max}=2$. In all other cases traffic jams are finite. We note that nevertheless, there can be a discontinuity in the mean velocity of cars ~\cite{NSp0transition2} or its derivative, reflecting a sudden onset of jams. This situation, where there is a discontinuous transition but no macroscopic condensate, appears different from traditional equilibrium transitions, and is likely related to the non-equilibrium nature of the system.

A wider applicability of the criterion is demonstrated by analyzing condensation in the Velocity Dependent Randomization model, in which the creation rate of new traffic jams is controlled by two stochastic parameters, one for the fluctuations of free flowing traffic ($p_f$), and one for those of jammed traffic ($p_j$). Exploring both parameters, we demonstrated a new condensate transition in the limit $p_f, p_j \rightarrow 0$ if $p_f$ vanishes faster than $p_j$, in agreement with the idea behind our criterion for condensation.

\section{Acknowledgements}
We thank B.Nienhuis for helpful discussions. This work was supported by the Complexity program of the Dutch Organization for Scientific Research (NWO). ASdW's contribution is supported by the Unga Forskare grant from the Swedish Research Council.

\bibliography{miedema_etal}

\begin{thebibliography}{25}
\expandafter\ifx\csname natexlab\endcsname\relax\def\natexlab#1{#1}\fi
\expandafter\ifx\csname bibnamefont\endcsname\relax
  \def\bibnamefont#1{#1}\fi
\expandafter\ifx\csname bibfnamefont\endcsname\relax
  \def\bibfnamefont#1{#1}\fi
\expandafter\ifx\csname citenamefont\endcsname\relax
  \def\citenamefont#1{#1}\fi
\expandafter\ifx\csname url\endcsname\relax
  \def\url#1{\texttt{#1}}\fi
\expandafter\ifx\csname urlprefix\endcsname\relax\def\urlprefix{URL }\fi
\providecommand{\bibinfo}[2]{#2}
\providecommand{\eprint}[2][]{\url{#2}}

\bibitem[{\citenamefont{Chowhury et~al.}(2000)\citenamefont{Chowhury, Santen,
  and Schadschneider}}]{Chowdhury_review}
\bibinfo{author}{\bibfnamefont{D.}~\bibnamefont{Chowhury}},
  \bibinfo{author}{\bibfnamefont{L.}~\bibnamefont{Santen}}, \bibnamefont{and}
  \bibinfo{author}{\bibfnamefont{A.}~\bibnamefont{Schadschneider}},
  \bibinfo{journal}{Phys.~Rep.} \textbf{\bibinfo{volume}{329}},
  \bibinfo{pages}{199} (\bibinfo{year}{2000}).

\bibitem[{\citenamefont{Bricard et~al.}(2013)\citenamefont{Bricard, Caussin,
  Desreurmaux, Dauchot, and Bartolo}}]{bartolo}
\bibinfo{author}{\bibfnamefont{A.}~\bibnamefont{Bricard}},
  \bibinfo{author}{\bibfnamefont{J.-B.} \bibnamefont{Caussin}},
  \bibinfo{author}{\bibfnamefont{N.}~\bibnamefont{Desreurmaux}},
  \bibinfo{author}{\bibfnamefont{O.}~\bibnamefont{Dauchot}}, \bibnamefont{and}
  \bibinfo{author}{\bibfnamefont{D.}~\bibnamefont{Bartolo}},
  \bibinfo{journal}{Nature} \textbf{\bibinfo{volume}{503}}, \bibinfo{pages}{95}
  (\bibinfo{year}{2013}).

\bibitem[{\citenamefont{van~der Weele et~al.}(2001)\citenamefont{van~der Weele,
  van~der Meer, Versluis, and Lohse}}]{granular_gas}
\bibinfo{author}{\bibfnamefont{K.}~\bibnamefont{van~der Weele}},
  \bibinfo{author}{\bibfnamefont{D.}~\bibnamefont{van~der Meer}},
  \bibinfo{author}{\bibfnamefont{M.}~\bibnamefont{Versluis}}, \bibnamefont{and}
  \bibinfo{author}{\bibfnamefont{D.}~\bibnamefont{Lohse}},
  \bibinfo{journal}{Europhys. Lett.} \textbf{\bibinfo{volume}{53}},
  \bibinfo{pages}{328} (\bibinfo{year}{2001}).

\bibitem[{\citenamefont{Evans et~al.}(2006)\citenamefont{Evans, Hanney, and
  Majumdar}}]{PFSS}
\bibinfo{author}{\bibfnamefont{M.~R.} \bibnamefont{Evans}},
  \bibinfo{author}{\bibfnamefont{T.}~\bibnamefont{Hanney}}, \bibnamefont{and}
  \bibinfo{author}{\bibfnamefont{S.~N.} \bibnamefont{Majumdar}},
  \bibinfo{journal}{Phys. Rev. Lett.} \textbf{\bibinfo{volume}{97}},
  \bibinfo{pages}{010602} (\bibinfo{year}{2006}).

\bibitem[{\citenamefont{Hirschberg et~al.}(2013)\citenamefont{Hirschberg,
  Mukamel, and Sch\"utz}}]{con_drift}
\bibinfo{author}{\bibfnamefont{O.}~\bibnamefont{Hirschberg}},
  \bibinfo{author}{\bibfnamefont{D.}~\bibnamefont{Mukamel}}, \bibnamefont{and}
  \bibinfo{author}{\bibfnamefont{G.~M.} \bibnamefont{Sch\"utz}},
  \bibinfo{journal}{Phys. Rev. E} \textbf{\bibinfo{volume}{87}},
  \bibinfo{pages}{052116} (\bibinfo{year}{2013}).

\bibitem[{\citenamefont{Majumdar et~al.}(2005)\citenamefont{Majumdar, Evans,
  and Zia}}]{ZRP_con}
\bibinfo{author}{\bibfnamefont{S.~N.} \bibnamefont{Majumdar}},
  \bibinfo{author}{\bibfnamefont{M.~R.} \bibnamefont{Evans}}, \bibnamefont{and}
  \bibinfo{author}{\bibfnamefont{R.~K.~P.} \bibnamefont{Zia}},
  \bibinfo{journal}{Phys. Rev. Lett.} \textbf{\bibinfo{volume}{94}},
  \bibinfo{pages}{180601} (\bibinfo{year}{2005}).

\bibitem[{\citenamefont{Spitzer}(1970)}]{ZRP}
\bibinfo{author}{\bibfnamefont{F.}~\bibnamefont{Spitzer}},
  \bibinfo{journal}{Adv. Math.} \textbf{\bibinfo{volume}{5}},
  \bibinfo{pages}{246} (\bibinfo{year}{1970}).

\bibitem[{\citenamefont{Evans and Hanney}(2005)}]{ZRP_review}
\bibinfo{author}{\bibfnamefont{M.~R.} \bibnamefont{Evans}} \bibnamefont{and}
  \bibinfo{author}{\bibfnamefont{T.}~\bibnamefont{Hanney}},
  \bibinfo{journal}{J.~Phys.~A: Math.~Gen.} \textbf{\bibinfo{volume}{38}},
  \bibinfo{pages}{195} (\bibinfo{year}{2005}).

\bibitem[{\citenamefont{Vigil et~al.}(1988)\citenamefont{Vigil, Ziff, and
  Lu}}]{chip1}
\bibinfo{author}{\bibfnamefont{D.}~\bibnamefont{Vigil}},
  \bibinfo{author}{\bibfnamefont{R.}~\bibnamefont{Ziff}}, \bibnamefont{and}
  \bibinfo{author}{\bibfnamefont{B.}~\bibnamefont{Lu}}, \bibinfo{journal}{Phys.
  Rev. B} \textbf{\bibinfo{volume}{38}}, \bibinfo{pages}{942}
  (\bibinfo{year}{1988}).

\bibitem[{\citenamefont{Rajesh and Krishnamurthy}(2002)}]{chip2}
\bibinfo{author}{\bibfnamefont{S.}~\bibnamefont{Rajesh}} \bibnamefont{and}
  \bibinfo{author}{\bibfnamefont{S.}~\bibnamefont{Krishnamurthy}},
  \bibinfo{journal}{Phys. Rev. E} \textbf{\bibinfo{volume}{66}},
  \bibinfo{pages}{046132} (\bibinfo{year}{2002}).

\bibitem[{\citenamefont{Barlovic et~al.}(2002)\citenamefont{Barlovic, Huisinga,
  Schadschneider, and Schreckenberg}}]{phase_sep_num}
\bibinfo{author}{\bibfnamefont{R.}~\bibnamefont{Barlovic}},
  \bibinfo{author}{\bibfnamefont{T.}~\bibnamefont{Huisinga}},
  \bibinfo{author}{\bibfnamefont{A.}~\bibnamefont{Schadschneider}},
  \bibnamefont{and}
  \bibinfo{author}{\bibfnamefont{M.}~\bibnamefont{Schreckenberg}},
  \bibinfo{journal}{Phys. Rev. E} \textbf{\bibinfo{volume}{66}},
  \bibinfo{pages}{046113} (\bibinfo{year}{2002}).

\bibitem[{\citenamefont{Kaupu{\v z}s et~al.}(2005)\citenamefont{Kaupu{\v z}s,
  Mahnke, and Harris}}]{traffic_zrp}
\bibinfo{author}{\bibfnamefont{J.}~\bibnamefont{Kaupu{\v z}s}},
  \bibinfo{author}{\bibfnamefont{R.}~\bibnamefont{Mahnke}}, \bibnamefont{and}
  \bibinfo{author}{\bibfnamefont{R.}~\bibnamefont{Harris}},
  \bibinfo{journal}{Phys. Rev. E} \textbf{\bibinfo{volume}{72}},
  \bibinfo{pages}{056125} (\bibinfo{year}{2005}).

\bibitem[{\citenamefont{Levine et~al.}(2004)\citenamefont{Levine, Ziv, and
  Mukamel}}]{traffic_chipping1}
\bibinfo{author}{\bibfnamefont{E.}~\bibnamefont{Levine}},
  \bibinfo{author}{\bibfnamefont{G.}~\bibnamefont{Ziv}}, \bibnamefont{and}
  \bibinfo{author}{\bibfnamefont{D.}~\bibnamefont{Mukamel}},
  \bibinfo{journal}{J.~Stat.~Phys.} \textbf{\bibinfo{volume}{117}},
  \bibinfo{pages}{819} (\bibinfo{year}{2004}).

\bibitem[{\citenamefont{Nagel and Schreckenberg}(1992)}]{NS}
\bibinfo{author}{\bibfnamefont{K.}~\bibnamefont{Nagel}} \bibnamefont{and}
  \bibinfo{author}{\bibfnamefont{M.}~\bibnamefont{Schreckenberg}},
  \bibinfo{journal}{J.~Phys.~I (France)} \textbf{\bibinfo{volume}{2}},
  \bibinfo{pages}{2221} (\bibinfo{year}{1992}).

\bibitem[{\citenamefont{Nagel}(1994)}]{lifetimes}
\bibinfo{author}{\bibfnamefont{K.}~\bibnamefont{Nagel}}, \bibinfo{journal}{Int.
  J. Mod. Phys. C} \textbf{\bibinfo{volume}{5}}, \bibinfo{pages}{567}
  (\bibinfo{year}{1994}).

\bibitem[{\citenamefont{Gerwinski and Krug}(1999)}]{gerwinski}
\bibinfo{author}{\bibfnamefont{M.}~\bibnamefont{Gerwinski}} \bibnamefont{and}
  \bibinfo{author}{\bibfnamefont{J.}~\bibnamefont{Krug}},
  \bibinfo{journal}{Phys. Rev. E} \textbf{\bibinfo{volume}{60}},
  \bibinfo{pages}{188} (\bibinfo{year}{1999}).

\bibitem[{\citenamefont{Roters et~al.}(1999)\citenamefont{Roters, L\"ubeck, and
  Usadel}}]{NSphasetransitionpolemic}
\bibinfo{author}{\bibfnamefont{L.}~\bibnamefont{Roters}},
  \bibinfo{author}{\bibfnamefont{S.}~\bibnamefont{L\"ubeck}}, \bibnamefont{and}
  \bibinfo{author}{\bibfnamefont{K.~D.} \bibnamefont{Usadel}},
  \bibinfo{journal}{Phys.~Rev.~E} \textbf{\bibinfo{volume}{59}},
  \bibinfo{pages}{2673} (\bibinfo{year}{1999}), \bibinfo{note}{and comments: D.
  Chowdhury, J. Kertesz, K. Nagel, L. Santen, A. Schadschneider, Phys.~Rev.~E.
  {\bf 61}, 3270 (2000), L. Roters, S. L\"ubeck, and K. D. Usadel, Phys. Rev. E
  {\bf 61}, 3272 (2000)}.

\bibitem[{\citenamefont{Boccara and Fuks}(2000)}]{Boccara}
\bibinfo{author}{\bibfnamefont{N.}~\bibnamefont{Boccara}} \bibnamefont{and}
  \bibinfo{author}{\bibfnamefont{H.}~\bibnamefont{Fuks}},
  \bibinfo{journal}{J.~Phys.~A: Math. Gen.} \textbf{\bibinfo{volume}{33}},
  \bibinfo{pages}{3407} (\bibinfo{year}{2000}).

\bibitem[{\citenamefont{Chen and Huang}(2001)}]{Chen}
\bibinfo{author}{\bibfnamefont{S.-P.} \bibnamefont{Chen}} \bibnamefont{and}
  \bibinfo{author}{\bibfnamefont{D.-W.} \bibnamefont{Huang}},
  \bibinfo{journal}{Phys.~Rev.~E} \textbf{\bibinfo{volume}{63}},
  \bibinfo{pages}{036110} (\bibinfo{year}{2001}).

\bibitem[{\citenamefont{Nagel and Paczuski}(1995)}]{NagelPaczuski}
\bibinfo{author}{\bibfnamefont{K.}~\bibnamefont{Nagel}} \bibnamefont{and}
  \bibinfo{author}{\bibfnamefont{M.}~\bibnamefont{Paczuski}},
  \bibinfo{journal}{Phys.~Rev.~E} \textbf{\bibinfo{volume}{51}},
  \bibinfo{pages}{2909} (\bibinfo{year}{1995}).

\bibitem[{\citenamefont{Eisenbl\"atter
  et~al.}(1998)\citenamefont{Eisenbl\"atter, Santen, Schadschneider, and
  Schreckenberg}}]{NSp0transition1}
\bibinfo{author}{\bibfnamefont{B.}~\bibnamefont{Eisenbl\"atter}},
  \bibinfo{author}{\bibfnamefont{L.}~\bibnamefont{Santen}},
  \bibinfo{author}{\bibfnamefont{A.}~\bibnamefont{Schadschneider}},
  \bibnamefont{and}
  \bibinfo{author}{\bibfnamefont{M.}~\bibnamefont{Schreckenberg}},
  \bibinfo{journal}{Phys.~Rev.~E} \textbf{\bibinfo{volume}{57}},
  \bibinfo{pages}{1309} (\bibinfo{year}{1998}).

\bibitem[{\citenamefont{Souza and Vilar}(2009)}]{NSp0transition2}
\bibinfo{author}{\bibfnamefont{A.~M.~C.} \bibnamefont{Souza}} \bibnamefont{and}
  \bibinfo{author}{\bibfnamefont{L.~C.~Q.} \bibnamefont{Vilar}},
  \bibinfo{journal}{Phys.~Rev.~E} \textbf{\bibinfo{volume}{80}},
  \bibinfo{pages}{021105} (\bibinfo{year}{2009}).

\bibitem[{\citenamefont{Schadschneider}(1999)}]{Schadschneider1999}
\bibinfo{author}{\bibfnamefont{A.}~\bibnamefont{Schadschneider}},
  \bibinfo{journal}{Eur.\ Phys.\ J.\ B\ -\ Cond.\ Mat.\ and Complex Systems}
  \textbf{\bibinfo{volume}{10}}, \bibinfo{pages}{573} (\bibinfo{year}{1999}).

\bibitem[{\citenamefont{Schreckenberg et~al.}(1995)\citenamefont{Schreckenberg,
  Schadschneider, Nagel, and Ito}}]{vmax}
\bibinfo{author}{\bibfnamefont{M.}~\bibnamefont{Schreckenberg}},
  \bibinfo{author}{\bibfnamefont{A.}~\bibnamefont{Schadschneider}},
  \bibinfo{author}{\bibfnamefont{K.}~\bibnamefont{Nagel}}, \bibnamefont{and}
  \bibinfo{author}{\bibfnamefont{N.}~\bibnamefont{Ito}},
  \bibinfo{journal}{Phys. Rev. E} \textbf{\bibinfo{volume}{51}},
  \bibinfo{pages}{2939} (\bibinfo{year}{1995}).

\bibitem[{\citenamefont{Barlovic et~al.}(2001)\citenamefont{Barlovic,
  Schadschneider, and Schreckenberg}}]{Barlovic}
\bibinfo{author}{\bibfnamefont{R.}~\bibnamefont{Barlovic}},
  \bibinfo{author}{\bibfnamefont{A.}~\bibnamefont{Schadschneider}},
  \bibnamefont{and}
  \bibinfo{author}{\bibfnamefont{M.}~\bibnamefont{Schreckenberg}},
  \bibinfo{journal}{J.~Phys.~A: Math.~Gen.} \textbf{\bibinfo{volume}{294}},
  \bibinfo{pages}{525} (\bibinfo{year}{2001}).

\end{thebibliography}

\end{document}